# Asymmetric soliton mobility in competing linear-nonlinear $\mathcal{PT}$-symmetric lattices

YAROSLAV V. KARTASHOV,[1,2,*] VICTOR A. VYSLOUKH,[3] AND LLUIS TORNER[1,4]

[1]*ICFO-Institut de Ciencies Fotoniques, The Barcelona Institute of Science and Technology, 08860 Castelldefels (Barcelona), Spain*
[2]*Institute of Spectroscopy, Russian Academy of Sciences, Troitsk, Moscow, 142190, Russian Federation*
[3]*Universidad de las Americas Puebla, Santa Catarina Martir, 72820, Puebla, Mexico*
[4]*Universitat Politecnica de Catalunya, 08034, Barcelona, Spain*
*Corresponding author: Yaroslav.Kartashov@icfo.eu

**We address the transverse mobility of spatial solitons in competing parity-time-symmetric linear and nonlinear lattices. The competition between out-of-phase linear and nonlinear lattices results in a drastic mobility enhancement within a range of soliton energies. We show that within such range, the addition of even a small imaginary part in the linear potential makes soliton mobility strongly asymmetric. The minimal phase tilt required for setting solitons into radiationless motion across the lattice in the direction opposite to that of the internal current drops to nearly zero, while the minimal phase tilt required for motion in the opposite direction notably increases. For a given initial phase tilt, the velocity of soliton motion grows with an increase of the balanced gain/losses. In this regime of enhanced mobility, tilted solitons can efficiently drag other solitons that were initially at rest, to form moving soliton pairs.**

Soliton mobility in discrete waveguide arrays and optical lattices is a topic of continuously renewed interest (see [1-4] and reviews [5,6]). When a shallow optical lattice is imprinted in a medium with uniform focusing Kerr nonlinearity, a minimal phase tilt is needed to set solitons into motion across the lattice–a manifestation of the so-called Peierls-Nabarro potential barrier [7]. In addition, solitons moving in periodically inhomogeneous media radiate energy away and may be captured in different lattice sites [1]. Various approaches have been suggested for the control and enhancement of the soliton mobility. They include optimization of the lattice profile [8] and utilization of quasi-periodic structures [9]. Mobility of high-power solitons was found to be enhanced in materials with saturable [10,11], quadratic [12] and competing [13] nonlinearities. In materials with a nonlocal nonlinearity that is asymmetric, such as diffusion nonlinearity in photorefractive crystals, solitons with sufficiently high energy start drifting across the lattice [14]. A mobility enhancement was also found in lattices with symmetric nonlocal nonlinearity [15]. An interesting approach to soliton mobility control relies on the utilization of materials with inhomogeneous nonlinearities, i.e., mixed linear-nonlinear lattices. Out-of-phase modulation of the linear and nonlinear refractive indices may lead to nearly radiationless propagation [16-20]. Mobility can be enhanced both for small-amplitude solitons extending over multiple lattice sites [17] and for high-amplitude states, whose width is smaller than the period of the potential [18].

All of the above-mentioned settings are conservative. New phenomena may occur in dissipative $\mathcal{PT}$-symmetric systems, where stationary evolution of linear and nonlinear excitations is possible due to a delicate balance between spatially modulated gain and losses (stationary solutions in such systems have internal currents) [21,22]. $\mathcal{PT}$-symmetry has been observed in optical systems [23,24]. In the presence of nonlinearity such systems support the formation of stable solitons whose properties have been studied in a number of works reviewed in [25,26]. In particular, solitons have been studied not only in usual linear $\mathcal{PT}$-symmetric lattices [27], but also in their nonlinear [28] and mixed [29] counterparts. However, the mobility of *spatial solitons* remains unexplored in continuous $\mathcal{PT}$-symmetric lattices. The existence of directions in such lattices associated with internal energy currents from domains with gain into domains with losses suggests the possibility of considerable asymmetry in mobility for solitons launched in opposite directions (by analogy with the asymmetric transport in different $\mathcal{PT}$-symmetric structures [30]).

In this Letter we study soliton mobility in competing linear $\mathcal{PT}$-symmetric and nonlinear lattices and we find that introduction of even a small imaginary part in the lattice induces a strong asymmetry in the mobility of solitons in the range where linear and nonlinear lattices nearly cancel each other. The asymmetry in mobility is a salient property of $\mathcal{PT}$-symmetric lattices - distinguishing them from conservative counterparts - and it is connected with considerable asymmetric variations of the amplitudes of solitons launched towards either the amplifying or the absorbing domains.

Light beam propagation in dissipative lattices with an out-of-phase modulation of the real part of the linear refractive index and the nonlinearity coefficient is governed by the nonlinear Schrödinger equation for the dimensionless field amplitude $q$:

$$i\frac{\partial q}{\partial \xi}=-\frac{1}{2}\frac{\partial^2 q}{\partial \eta^2}-[1-\sigma\cos^2(\Omega\eta)]q|q|^2- \quad (1)$$
$$[p_{\rm re}\cos^2(\Omega\eta)-ip_{\rm im}\sin(2\Omega\eta)]q,$$

where $\eta$ and $\xi$ are the normalized transverse and longitudinal coordinates, respectively; $2\Omega$ is the spatial frequency of all lattices, including the linear and nonlinear ones; $\sigma$ is the nonlinearity modulation depth (see [20] for description of settings with inhomogeneous nonlinearity landscapes); $p_{\rm re}$ is proportional to the modulation depth of the refractive index; and $p_{\rm im}$ is the strength of the gain/losses in the system. In the linear case, the solutions of Eq. (1) are Bloch modes. The eigenvalue spectrum remains real as long as $p_{\rm im}<p_{\rm re}/2$. Therefore, we will set $p_{\rm im}$ values that are substantially lower than those that correspond to the $\mathcal{PT}$-symmetry breaking threshold $p_{\rm im}=p_{\rm re}/2$ [22]. The nonlinear lattice $1-\sigma\cos^2(\Omega\eta)$ is out-of-phase with the real part $p_{\rm re}\cos^2(\Omega\eta)$ of the linear lattice. Therefore, they compete because self-focusing acts toward concentration of light in the regions where nonlinearity is higher, but the linear refractive index is reduced [18]. The impact of the nonlinear lattice becomes important at high peak amplitudes and is negligible when $q\to 0$. Current technologies allow relatively deep simultaneous modulation of the linear refractive index and the nonlinearity coefficient of the material [20], as it occurs for example upon fs-laser writing of waveguide arrays in various materials [31]. Combined with inhomogeneous doping with active centers and pumping, this technique may allow fabricating mixed dissipative lattices. For a characteristic scale $x_0\sim 20~\mu\text{m}$ at the wavelength $\lambda=0.8~\mu\text{m}$, the diffraction length amounts to about $kx_0^2\sim 4.6$ mm and $\Omega=2$ corresponds to a lattice period of $31~\mu\text{m}$. A lattice depth $p_{\rm re}=4$ corresponds to a refractive index contrast $\delta n_{\rm re}\sim 10^{-4}$, while $p_{\rm im}=0.4$ corresponds to an amplification/absorption length of about $11.5$ mm. Assuming a nonlinear coefficient of $n_2\simeq 2.7\times 10^{-16}~\text{cm}^2/\text{W}$ a dimensionless intensity $|q|^2=1$ corresponds to some $10^{11}~\text{W}/\text{cm}^2$. All effects discussed here are observable at ten diffraction lengths, i.e. at $\sim 46$ mm.

When $\sigma, p_{\rm re}, p_{\rm im}\ll 1$ linear and nonlinear lattices act as small perturbations of a classical cubic Schrödinger equation, whose single-soliton solution $q_s=\kappa\,\text{sech}[\kappa(\eta-\eta_c)-V\xi]\exp(i\phi)$ may serve as a zero-order approximation for the analysis of soliton mobility in a shallow lattice. Here $\kappa$ is the form-factor, $V$ is the transverse soliton velocity, and $\eta_c$ is the deviation of soliton trajectory from a straight line; $\phi(\eta,\xi)=-(V^2-\kappa^2)\xi/2+V(\eta-\eta_c)-\phi_0$ is the phase. We use an approach based on the inverse scattering transform [32] to analyze the perturbations of the soliton parameters induced by shallow lattices that are described by:

$$\delta\kappa=\int_{-\infty}^{\infty}\kappa\,\text{sech}(\kappa\eta)\text{Re}[\exp(iV\eta)\delta q]d\eta,$$
$$\delta V=\int_{-\infty}^{\infty}\text{sech}(\kappa\eta)\tanh(\kappa\eta)\,\text{Im}[\exp(iV\eta)\delta q]d\eta, \quad (2)$$
$$\delta\eta_c=\int_{-\infty}^{\infty}\eta\,\text{sech}(\kappa\eta)\text{Re}[\exp(iV\eta)\delta q]d\eta,$$
$$\delta\phi_0=\int_{-\infty}^{\infty}\text{sech}(\kappa\eta)[1-\kappa\eta\tanh(\kappa\eta)]\text{Im}[\exp(iV\eta)\delta q]d\eta.$$

Here $\delta q$ is the perturbation of single soliton solution $q_s(\eta,\xi)$ accumulated upon propagation over an infinitesimal distance $d\xi$:

$$\delta q=[-i\sigma\cos^2(\Omega\eta)|q_s|^2+ip_{\rm re}\cos^2(\Omega\eta)+ \quad (3)$$
$$p_{\rm im}\sin(2\Omega\eta)]q_s d\xi.$$

To simplify the calculation of the integrals in (2) one can consider a mathematically equivalent problem when an initially immobile soliton is dragged by moving (in the opposite direction) lattice $\sim\cos^2[\Omega(\eta+V\xi)], \sin[2\Omega(\eta+V\xi)]$. Integration (performed also with respect to $\xi$) predicts that the mixed lattice will induce the following *total* variation of the soliton velocity at distance $\xi$:

$$\delta V(\xi)=[\sigma(\kappa^2+\Omega^2)/3-p_{\rm re}]F_S(\pi\Omega/\kappa)\sin^2(2V\Omega\xi)/V, (4)$$

where the function $F_S(x)=x/\sinh(x)$ monotonically decreases to zero as $x\to\infty$. This means that the impact of the lattice on soliton velocity reduces with the decrease of the form-factor $\kappa$ or increase of the lattice frequency $\Omega$. $\delta V$ also decreases with the increase of velocity $V$. Importantly, refractive index modulation slows down the soliton launched at $\eta=0$, while the nonlinear lattice accelerates it. Exact cancellation of impact of two lattices occurs for particular form-factor $\kappa=(3p_{\rm re}/\sigma-\Omega^2)^{1/2}$, at which $\delta V=0$ and perfect mobility is predicted. At the same time, the imaginary part of the lattice is responsible for total periodic variations of the form-factor:

$$\delta\kappa(\xi)=2p_{\rm im}\kappa(0)F_S(\pi\Omega/\kappa)\sin^2(2V\Omega\xi)/V\Omega \quad (5)$$

and, consequently, of its energy flow $U(\xi)=\int_{-\infty}^{\infty}|q|^2\,d\eta=2\kappa(\xi)$. The amplitude of these oscillations is maximal for strongly localized states with $\kappa\gg 1$ and diminishes with growing velocity $V$. It should be stressed that the sign of $\delta\kappa$ changes upon inversion of the velocity sign $V$: it is *positive* for $V>0$ and *negative* for $V<0$. It is this effect combined with notable velocity oscillations that leads to *asymmetry* in the mobility for the right- and left-launched solitons illustrated below. The effective modification of the form-factor drives solitons into or outside the domain with enhanced mobility.

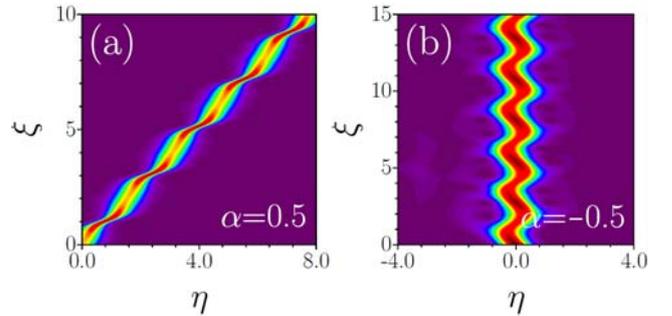

Fig. 1. (Color online) Propagation of solitons with opposite phase tilts in a $\mathcal{PT}$-symmetric lattice with $p_{\rm im}=0.4$, $U=11$, $\sigma=0.4$, $p_{\rm re}=4$.

Gain/losses also lead to additional dynamical shifts of the soliton center upon propagation, whose rate is described by

$$\delta\eta_c/d\xi=p_{\rm im}F_S(\pi\Omega/\kappa)F_C(\pi\Omega/\kappa)\cos(2V\Omega\xi)/\Omega, \quad (6)$$

where the function $F_C(x)=x\coth(x)-1$. Such periodically oscillating shift contributes to the soliton velocity. The amplitude of the soliton center oscillations grows from zero upon increase of $\pi\Omega/\kappa$, reaches its maximum for a certain $\pi\Omega/\kappa$ value, and asymptotically approaches zero with further growth of this parameter.

To extend the analytical results obtained for shallow lattices to the case of lattices of arbitrary depth, we first found exact soliton solutions of Eq. (1) with competing lattices in the form

$q_s(\eta, \xi) = w(\eta)e^{ib\xi}$, where $b$ is the propagation constant and $w(\eta)$ is a function (which is complex due to the presence of internal currents required for stationary propagation at $p_{\text{im}} \neq 0$) describing the soliton profile. When $p_{\text{im}}$ is relatively small, solitons in a $\mathcal{PT}$-symmetric lattice exhibit an almost linear internal phase tilt. In order to set such stationary states into motion across the lattice we impose on them the additional input phase tilt $\exp(i\alpha\eta)$ and let them propagate. We set the lattice frequency $\Omega = 2$ and consider sufficiently deep structures with $p_{\text{re}} = 4$, unless stated otherwise. It was shown previously for conservative lattices [18] that the soliton mobility is strongly enhanced in the high-amplitude limit, where the linear and nonlinear lattices compete on similar footing. We thus select the energy flow $U \sim 11$ so that solitons fall close to the point where the critical tilt required for setting them into motion in conservative lattices drops to zero. Namely, around such point inclusion of the imaginary part of the potential drastically changes the soliton dynamics, because oscillations of the energy flow induced by the imaginary part of the lattice may easily shift soliton into the mobility domain or vice versa, into the domain where excitations are pinned.

This behavior is illustrated in Fig. 1 where we compare the evolution of the same input, but for opposite phase tilts $\alpha = 0.5$ and $\alpha = -0.5$. While in the former case the soliton starts moving across the lattice with considerable velocity, in the latter case it cannot leave the input channel and oscillates inside it. In accordance with the predictions of the perturbation approach, solitons accelerate towards the amplifying domains, where their amplitude grows, and slow down upon passage of absorbing domains, where their amplitude decreases. Although $p_{\text{im}} = 0.4$ is small, the oscillations of the soliton peak-amplitude are much larger than in the conservative lattice. This effect arises because gain/losses affect the soliton form-factor already at first-order perturbation theory [Eq. (5)], while the conservative linear and nonlinear lattices can affect it only at second-order.

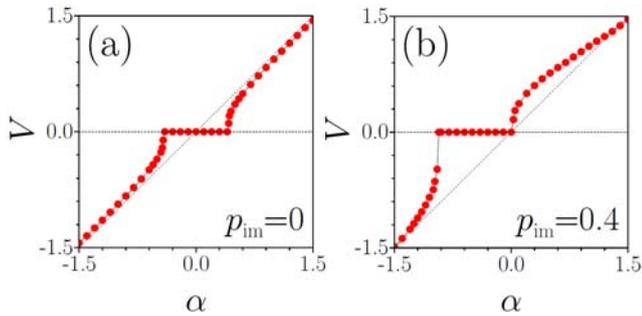

Fig. 2. (Color online) Velocity of solitons escaping the central channel versus the phase tilt $\alpha$ for $p_{\text{im}} = 0$ (a) and $p_{\text{im}} = 0.4$ (b). In both cases $\sigma = 0.4$, $p_{\text{re}} = 4$. The diagonal dashed line corresponds to $V = \alpha$.

Therefore, inclusion of the imaginary part of the lattice may result in strong *asymmetry* of the mobility, which is the central result of this Letter. This is illustrated in Fig. 2, where we compare the dependencies of the output distance-averaged soliton velocity $V$ on the input phase tilt $\alpha$ in the conservative [Fig. 2(a)] and in the $\mathcal{PT}$-symmetric [Fig. 2(b)] lattices. In both cases there is a range of tilts where solitons remain immobile (i.e., the tilt is insufficient to overcome the potential barrier). In the conservative lattice, the $V(\alpha)$ dependence is exactly antisymmetric, indicating that there is no difference in dynamics for left-launched and right-launched solitons. The output velocity gradually approaches $\alpha$ when $\alpha \to \pm\infty$. In complete contrast, in the $\mathcal{PT}$-symmetric lattice the dependence $V(\alpha)$ is strongly *asymmetric*: the minimal phase tilt required to set soliton into motion in the right direction drops to nearly zero, but the $\alpha$ value needed to force soliton to travel to the left increases by nearly a factor of three. Moreover, the velocity of solitons moving to the right may now *exceed* the initial phase tilt. As mentioned above, this enhanced or reduced mobility originates in the oscillations of the soliton amplitude and instantaneous velocity even at $p_{\text{im}} \ll p_{\text{re}}$: one can clearly see that at the initial stages of propagation the amplitude and velocity grow for a right-launched state (thereby shifting it into domain with enhanced mobility) and decrease for a left-launched one (thereby shifting soliton into domain where excitations are pinned).

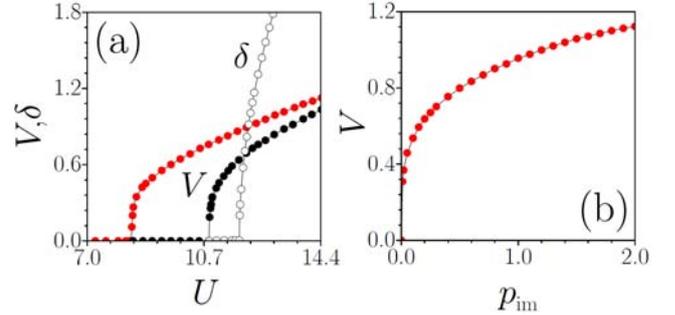

Fig. 3. (Color online) (a) Soliton velocity versus $U$ for $p_{\text{im}} = 0$ (black circles), $p_{\text{im}} = 0.4$ (red circles) for $\alpha = 0.5$ and perturbation growth rate for stationary states versus $U$ for $p_{\text{im}} = 0.4$ (open circles). (b) $V$ versus $p_{\text{im}}$ for $\alpha = 0.5$, $U = 10.9$. In all cases $p_{\text{re}} = 4$, $\sigma = 0.4$.

The velocity of soliton motion at fixed $\alpha$ can be controlled by modifying the energy flow [Fig. 3(a)]. In both conservative (black circles) and dissipative (red circles) settings, solitons start moving when $U$ exceeds a certain minimal value. In the $\mathcal{PT}$-symmetric case this value is substantially lower than in the conservative case, despite the fact that the imaginary part of the lattice is rather small. In [18] it was conjectured that the mobility enhancement occurs close to the point where odd solitons lose their stability. The line with open circles in Fig. 3(a) illustrates the dependence of the growth rate on $U$ for the most unstable perturbation mode of such solitons in the $\mathcal{PT}$-symmetric lattice. While the border of the stability domain is nearly the same for $p_{\text{im}} = 0$ and $p_{\text{im}} = 0.4$, the border of the *mobility* domain differs substantially for these two $p_{\text{im}}$ values.

The velocity of motion of right-launched solitons can be controlled *exclusively* by the imaginary part of the linear lattice. In Fig. 3(c) we show the $V(p_{\text{im}})$ dependence up to the symmetry-breaking point. The soliton velocity monotonically increases with $p_{\text{im}}$.

Finally, we study the impact of the asymmetrically-enhanced soliton mobility in the case of moving soliton pairs. To excite the pairs we use at the input two identical (in-phase) solitons $q|_{\xi=0} = q_s(\eta)\exp(i\alpha\eta) + q_s(\eta + d)$ residing in adjacent lattice channels (here $d = \pi/\Omega$), and we impose an initial phase tilt $\alpha$ on only one of them (the one residing in the right-hand-side). Then, the soliton located at the left is initially immobile, while the one located at the right is launched with some positive velocity. If $\alpha$ is small the attraction between solitons results in their collision, after which the soliton that was initially located at the left is ejected from the initial pair of channels and starts moving with considerable velocity [Fig. 4(a)], while the soliton that was initially located at the right, gets trapped due to the existing strong mobility asymmetry. If $\alpha$ is too large the collision between the two solitons does not occur. Due to the enhanced mobility, even a relatively weak attractive force between two in-phase solitons is sufficient for ejection of the soliton that was initially at rest [Fig. 4(c)], after which two solitons propagate with substantially different velocities. By adjusting the phase tilt imposed on the dragging soliton one can achieve a situation when both solitons start moving across the lattice glued together. This regime occurring at specific tilt $\alpha = \alpha_C$ is illustrated in Fig. 4(b).

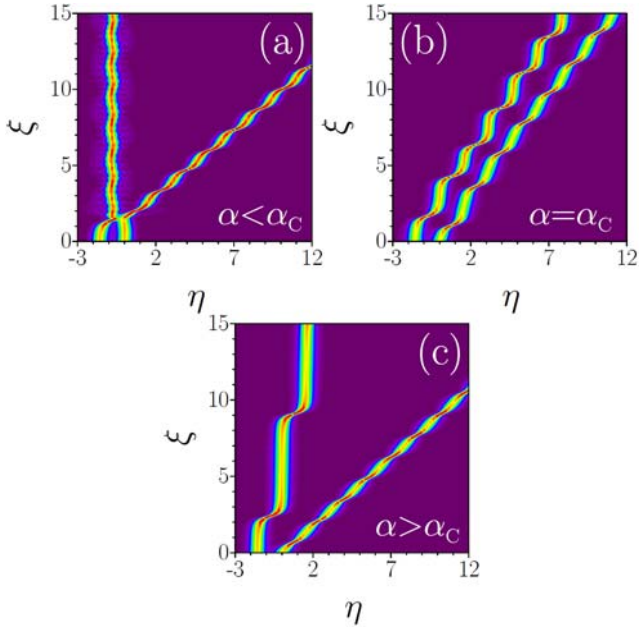

Fig. 4. (Color online) Interaction of two in-phase solitons with $U=12$ in the presence of initial phase tilt $\alpha = 0.15 < \alpha_C$ (a), $\alpha = 0.4006 \approx \alpha_C$ (b), $\alpha = 0.9 > \alpha_C$ (c) imposed on right soliton at $\sigma = 0.4$, $p_{re} = 4$, $p_{im} = 0.4$.

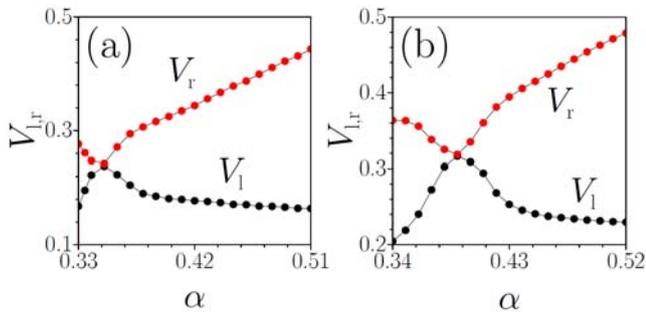

Fig. 5. (Color online) Velocities of the solitons located at the right and the left in the interacting pair versus the phase tilt $\alpha$ imposed on the soliton located at the right, for $U=10.9$ and (a) $p_{re} = 0.5$, $p_{im} = 0.05$, $\sigma = 0.05$, (b) $p_{re} = 1$, $p_{im} = 0.1$, $\sigma = 0.1$.

Figure 5 shows the dependencies of the output soliton velocities on the initial phase tilt $\alpha$ imposed on right soliton for lattices of different depths. In both cases shown [panels (a) and (b)], the parameters of the lattice and input solitons were adjusted in order to be close to the point where mobility enhancement occurs. As readily visible, the point where the two curves corresponding to different solitons touch each other (i.e. where solitons propagate as a single complex) shifts to larger values of $\alpha$ with the increase in lattice depth. Notice that eventually the pairs slowly separate because of the unavoidable presence of small, but non-negligible radiative losses.

Summarizing, we studied soliton mobility in $\mathcal{PT}$-symmetric lattices imprinted in media with inhomogeneous nonlinearity. The salient results reported is that, under appropriate conditions, the presence of even small gain/losses may substantially impact the soliton mobility by introducing a strong mobility *asymmetry*, which has been shown to affect both single solitons and soliton pairs.

This work has been partially supported by the Severo Ochoa Excellence program and by Fundació Cellex.